\documentclass[prl,amsmath,amssymb,reprint,onecolumn]{revtex4-1}
\usepackage{CJK}
\usepackage{graphicx}
\usepackage{dcolumn}
\usepackage{bm}
\usepackage[mathlines]{lineno}

\usepackage[utf8]{inputenc}
\usepackage[T1]{fontenc}
\usepackage{mathptmx} 

\usepackage{xcolor} 

\begin{document}
	
	\title{Intrinsic permeability of heterogeneous porous media}
	
	\author{Wenqiao Jiao}
	\affiliation{Institute of Earth Science, University of Lausanne, 1015 Lausanne, Switzerland} 
	\affiliation{Dipartimento di Ingegneria Civile e Ambientale, Politecnico di Milano, Piazza L. Da Vinci 32, 20133 Milano, Italy}
	\author{David Scheidweiler}
	\affiliation{Institute of Earth Science, University of Lausanne, 1015 Lausanne, Switzerland} 
	\author{Nolwenn Delouche}
	\affiliation{Institute of Earth Science, University of Lausanne, 1015 Lausanne, Switzerland} 
	\author{Alberto Guadagnini}
	\affiliation{Dipartimento di Ingegneria Civile e Ambientale, Politecnico di Milano, Piazza L. Da Vinci 32, 20133 Milano, Italy}
	\email[Corresponding author: ]{alberto.guadagnini@polimi.it}
	\affiliation{Department of Hydrology and Atmospheric Sciences, The University of Arizona, Tucson, AZ 85721, USA}
	\author{Pietro de Anna}
	\email[Corresponding author: ]{pietro.deanna@unil.ch}
	\affiliation{Institute of Earth Science, University of Lausanne, 1015 Lausanne, Switzerland} 
	\date{\today}
	
	\begin{abstract}
		\noindent
		Providing a sound appraisal of the nature of the relationship between flow $(Q)$ and pressure drop $(\Delta P)$ for porous media is a long-standing fundamental research challenge. A wide variety of environmental, societal and industrial issues, ranging, e.g., from water-soil system remediation to subsurface energy optimization, is affected by this critical issue. While such dependence is well represented by the Kozeny-Carman formulation for homogeneous media, the fundamental nature of such a relationship ($Q$ vs $\Delta P$) within heterogeneous porous systems characterized by a broad range of pore sizes is still not fully understood. We design a set of controlled and complex porous structures and quantify their intrinsic permeability through detailed high quality microfluidics experiments. We synthesize the results upon deriving an original analytical formulation relating the overall intrinsic permeability of the porous structure and their key features. Our formulation explicitly embeds the spatial variability of pore sizes into the medium permeability through a conceptualization of the system as a collection of smaller scale porous media arranged in series. The resulting analytical formulation yields permeability values matching their experimentally-based counterparts without the need of additional tunable parameters. Our study then documents and supports the strong role played by the micro-structure on the overall medium permeability. \\
		
	\end{abstract}
	
	\maketitle
	
	\section{Introduction}
	
	\noindent
	Characterization of fluid flow through permeable media is critical in a wide variety of natural and engineered scenarios. These include, for example, riverbank filtration~\cite{banzhaf2011investigative}, remediation of contaminated soil~\cite{qiu2021experiments}, deep geothermal energy exploration and production~\cite{watanabe2020stabilizing}, nucleation and recurrence of earthquakes~\cite{giger2007permeability}, and geological $CO_2$ storage~\cite{de2016influence}. The structural properties of porous systems that are known to control macroscopic fluid flow, as well as other processes related to mixing of chemicals, include porosity, grain size and shape, tortuosity, effective pore radius, solid surface areas, and pore/grain size distribution~\cite{graczyk2020predicting, luo2014numerical, liu2019pore, sperry1995model, cormican2020grain, kang2014pore}. \\
	
	\noindent
	In presence of a laminar regime associated with a flow rate $Q$ of a single fluid of dynamic viscosity $\mu$ through a porous domain of cross-sectional area $A$, a continuum (i.e., macroscopic) description of the average fluid velocity $q$ across the system is provided by Darcy's law~\cite{darcy1856fontaines,bear1988dynamics}
	\begin{equation}\label{eq p_Darcy}
		q = \frac{Q}{A} = - \frac{k}{\mu} \, \nabla P.
	\end{equation}
	Here, $k$ denotes the intrinsic medium permeability and $\nabla P = \Delta P/L$ is the overall pressure gradient (evaluated along the mean flow direction) induced by a pressure drop $\Delta P$ across the domain of size $L$. The intrinsic permeability $k$ represents the overall ability of the porous system to host a net flow and it is regarded as a key parameter for the porous medium characterization at a continuum (or Darcy) scale~\cite{drummond1984laminar, wagner2021permeability}.\\
	
	\noindent
	Providing a sound assessment of porous media permeability is still the object of intense research activities. Remarkable efforts have been devoted to explore correlations between permeability and the porous medium internal architecture, as quantified through quantities such as grain and/or pore size distributions, characteristic lengths or porosity~\cite{koponen1997permeability, sutera1993history, sullivan1942permeability, yang2014analytical, nishiyama2017permeability, schulz2019beyond, nomura2018modified,Dvorkin,hommel2018porosity}. Among these studies, the classical Kozeny-Carman formula~\cite{kozeny1927kapillare,carman1937fluid} is broadly employed. Such a formulation is based on a conceptual picture according to which the average (Darcy) velocity through a homogeneous porous medium can be expressed as the corresponding velocity associated with a system composed of a pipes collection with identical diameter ($d$). Flow within each pipe is, then, expressed through the Hagen-Poiseuille law, which is the exact solution of Stokes equations, representing momentum conservation, for laminar fluid displacement. Within this theoretical formulation, the intrinsic permeability $k$ results to be a function of the medium porosity $\phi$, as
	\begin{equation}\label{eqKC}
		k = \frac{c_0}{ \sigma^2} \, \frac{\phi^3}{(1 - \phi)^2},
	\end{equation}
	where $c_0 = \frac{1}{5}$ is the Kozeny constant and $\sigma$ denotes the specific surface of the medium grains~\cite{kozeny1927kapillare}. The value of $\sigma$ can be found invoking that the viscous friction experienced by the flowing fluid at the grain walls in the porous medium coincides with its counterpart at the pipe walls. Thus, for a homogeneous porous domain composed by an assemblage of identical spheres, of diameter $d_g$, $\sigma = \frac{6}{d_g}$, that honors the surface to volume ratio between spheres (the porous medium grains) and pipes (the model). This formlatio has been adapted to other homogeneous structures: for instance, considering an assemblage of vertical cylinders as grains, $\sigma = \frac{4}{d_g}$~\cite{schulz2019beyond, skartsis1992resin}. \\ 
	
	\noindent
	A large number of diverse porous systems have an internal structure that is characterized by a complex arrangement of pores. These, in turn, cannot be properly represented through an equivalent spatially homogeneous model of the kind described above and cannot be effectively characterized through a unique (averaged/characteristic) pore/grain size and porosity. Natural porous media are documented to possess a very rich structure in terms of pore and/or throat length distributions (e.g.~\cite{blunt13-awr}). For instance, a log-normal function has been found to represent the distribution of pore sizes across glass beads and sand porous materials~\cite{Gueven2017,Kazutaka2010}. Further analyses, performed on clay and sandstone samples through high-resolution imaging technologies (such as BIB-SEM, X-ray $\mu$-CT and FIB-SEM image-stacks)~\cite{HemesMM2015} or using small-angle neutron scattering and fluid-invasion methods~\cite{ZhaoSrep2017}, document power-law pore size distributions, thus highlighting the multi-scale nature of such permeable structures.\\
	
	\noindent
	Detailed characterizations of complex pore spaces are typically grounded on high-resolution imaging of a given sample, segmentation of the acquired images to identify pores and grains, and ensuing construction of a pore network. The latter step is often performed with methods based on Delaunay Triangulation or Maximum Inscribed Circles (MICs) algorithms~\cite{nguyen2021new, gostick2013random, gao2012two,liu2023random}. While values of average pore size and connectivity have been used to quantify the overall, macroscopic, medium permeability~\cite{muljadi2016impact}, the spatial variability of pore sizes, quantified by their statistical distribution, has been documented to control the local velocity values distribution  itself and, hence, asymptotic transport~\cite{deAnnaPRF2017}. Moreover, also the spatial correlation of pore sizes has been shown to affect flow and its patterns in porous media~\cite{alim2017local}.\\
	
	\noindent
	As pore sizes, $\lambda$, of natural porous media span across several orders of magnitude, a simplified picture that reduces the complexity of a porous structure to the one of a homogeneous assemblages of identical spheres, is expected to fail in representing heterogeneous systems. Alternative formulations, based on modifications of the Kozeny-Carman formula, propose different relationships between the average pore throat size, effective grain size and medium porosity. Key shortcomings of these models are that these still rely on the use of a single (average/representative) pore throat and/or grain size and that some parameters embedded therein have only minimal (or no) direct connection to the host medium structure and cannot be quantified through direct observations~\cite{xu2008developing,hyman2013pedotransfer}.\\
	
	\noindent
	Advancements in the development of pore network models~\cite{raoof2010new,liu2022pore,xiong2016review} to assess the macroscopic mass transport properties in porous media suggest that the effect of pore connectivity is markedly stronger than the effect of porosity~\cite{jivkov2013novel}. Moreover, the pore space structure of porous systems can be significantly altered over short time scales by a variety of phenomena such as, e.g., biofilm growth~\cite{shin2021coupled}, mineral dissolution~\cite{beckingham2017evaluation, balucan2016acid, ling2022probing} and/or precipitation~\cite{wu20193d,cuthbert2013field}.\\
		
	\noindent
	For practical reasons, porous samples are typically characterized at scales of a few tens of pores, at best. At this scale, an heterogeneous structure typically result in flow focused in high velocity channels and stagnation zones~\cite{deAnnaPRF2017}: a fluid particle entering a fast flow channel will probably never experience low velocities before leaving the medium or, in contrast, can enter the system in a stagnation zone within which it moves very slowly. The description of flow in these scenarios is controlled by the connection between pores, quantified by the system connectivity \cite{bernabe2011pore,davudov2018impact,davudov2020interplay}. However, here we consider the scenario of a heterogeneous porous system of size $L$ that is much larger than the average pore size. This scenario virtually corresponds to consider the possibility of scanning a sample across several length-scales (i.e., not just a few tens but several hundreds of average pore sizes). Hence, we hypothesize that each fluid parcel flowing through the porous medium experiences the whole structural variability therein and not only the one of a high velocity channel or a stagnation zone. In other words, we assume that the flow statistics along each trajectory is the same as the ensemble velocity statistics, i.e., the system under consideration is ergodic. \\
	
	\noindent
	To study such complex media, we design controlled and complex porous structures whose size $L$ is a few hundreds of average pore sizes. These are, then, printed into microfluidics devices that we characterize in terms of their overall permeability $k$, in presence of laminar flow driven by an imposed macroscopic pressure drop. We show that the ensuing permeability values cannot be interpreted through classical models of the kind described above and based on the use of effective and/or averaged quantities (e.g., representative grain size or porosity). Thus, we derive and test a novel analytical formulation that explicitly embeds the whole range of pore sizes to predict the system permeability. Our model predictions display a remarkable agreement with experimental data without the need of any tunable parameter. As such, our formulation provides a theoretically sound link between a heterogeneous porous medium structure and its permeability.	\\
	
	\begin{figure}[h!]
		\begin{center}
			\includegraphics[trim={0 5 5 0},clip,width=1\textwidth]{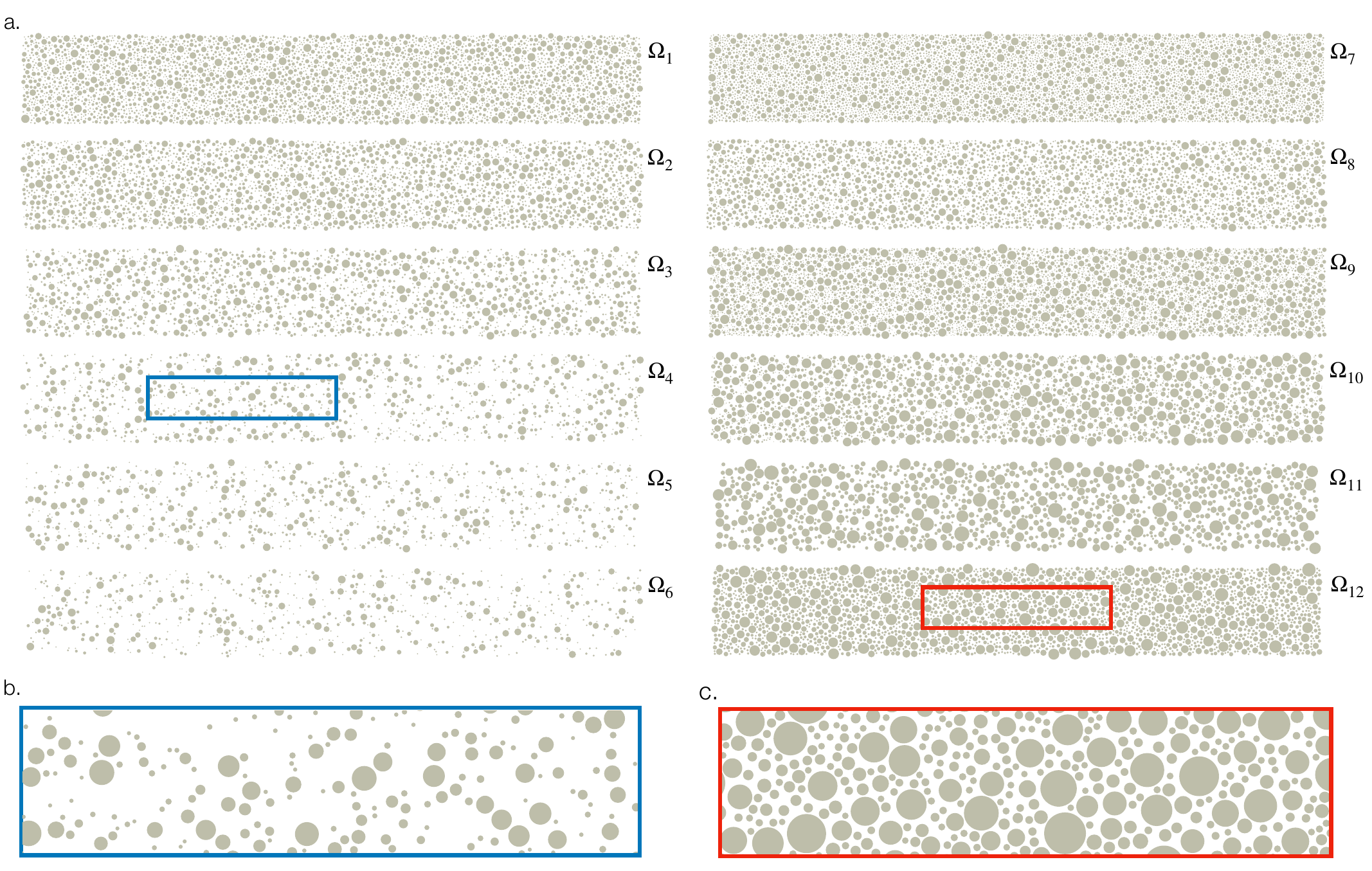} \\
			\caption{(a) Representation of the 12 two-dimensional porous geometries $\Omega_i$ (Length = $54.5$~mm, Width = $3.8$~mm) designed in this study, showing the non-overlapping circular disks (gray circles) of random position and radius $r_j$ associated with each geometry. eZoomed-in views of selected regions, i.e., (b) blue corresponding to medium $\Omega_4$ (porosity of 82.6\%) and (c) red corresponding to medium $ \Omega_{12}$ (porosity of 46.0\%).} \label{fig1}
		\end{center}
	\end{figure}
	
	\section{Porous media structure design}\label{sec2}
	
	\noindent
	To investigate the link between a heterogeneous porous structure and its intrinsic permeability $k$, we design 12 two-dimensional porous media whose structure ($\Omega_i \, (i = 1, …, 12)$) is composed by $N_d$ non-overlapping, circular and impermeable grains, here represented by $N_d$ circular disks of random position and radius $r_j$, for $j = 1, …, N_d$. Figure \ref{fig1} shows the 12 porous structures considered in this study. Single phase fluid flow takes place among the circular grains, i.e. the pore space, under the action of an imposed pressure gradient. \\
	
\begin{figure}[h!]
		\begin{center}
			\includegraphics[trim={0 0 0 0},clip,width=1\textwidth]{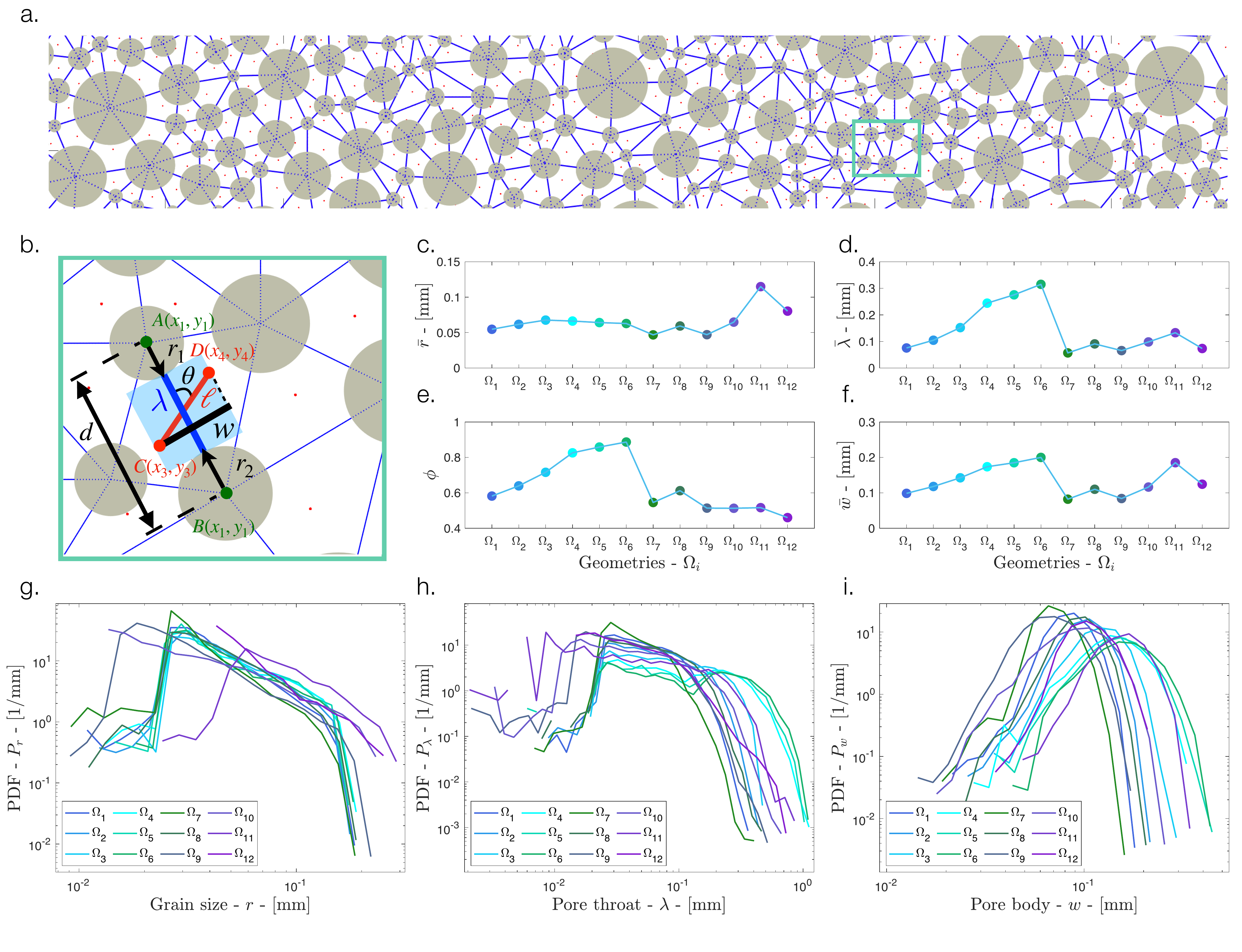} \\
			\caption{(a) Details of the Delaunay triangulation of a portion of a pore geometry (edges are marked in blue and their centers with red points). (b) Zoomed-in view of green rectangular in (a). Schematic of the conceptual model of pores as pipes (cyan squares) associated with pore throats $\lambda$ and pore bodies $w$. The blue solid line defines a pore throat of size $\lambda = d - r_1 - r_2$; each couple of neighboring triangle of center $C$ and $D$ defines a pore body $w = \ell ~ sin\theta$, where $\theta=\arccos\frac{ \vec{AB} \cdot \vec{CD} }{\lvert \vec{AB} \rvert \cdot \lvert \vec{CD} \rvert} \in [0,\pi]$. We set that angle to $\pi -\theta$ when it is greater than $\pi/2$. (c) Average grain radius $\bar{r}$ for each geometry $\Omega_i$. (d) Average pore throat $\bar{\lambda}$ for each geometry $\Omega_i$. (e) Porosity $\phi$ of each geometry $\Omega_i$. (f) Average pore body $\bar{w}$ for each geometry $\Omega_i$. (g) Double logarithmic plot of the probability density function (PDF), $P_r$, of the grain size $r$ distribution. (h) Double logarithmic plot of the pore throat size $\lambda$ PDF, $P_{\lambda}$. (i) Double logarithmic plot of the pore body size $w$ PDF, $P_{w}$.} \label{fig2}
		\end{center}
	\end{figure}

	\noindent
	We characterize each porous structure by identifying the nearest neighbors of each grain with a Delaunay triangulation associated with the grain centers, as schematically shown in Figure~\ref{fig2} (a,b). Each edge of the triangulation, of length $d$, connects the centers of two nearest neighboring grains centers of radii $r_1$ and $r_2$. This defines a pore throat of size $\lambda= d - r_1 - r_2$ and each triangle (i.e., a triplet of nearest neighbors) defines a pore body whose center is represented by a red dot in Figure~\ref{fig2}(a,b). The size $w$ of a pore body is the length over which the fluid flows while squeezed within a pore throat $w = \ell ~ sin\theta$, where $\ell$ is a segment measuring the distance between neighboring pore body centers and $\theta$ is the angle between $\ell$ (red segment in Figure~\ref{fig2}~b) and the pore throat $\lambda$ (blue segment in Figure~\ref{fig2}~b). \\
	
	\noindent
	We note that the media structures are designed to have similar macroscopic quantities: average grain radius $\overline{r}$, pore throat size $\overline{\lambda}$, pore body size $\overline{w}$ and porosity $\phi$, as shown in Figure~\ref{fig2}~(c-d-f-e), respectively. These macroscopic values vary at best by a factor of 3, across all porous structures. We further note that grain sizes, pore throats and bodies are also designed to be distributed as broadly as possible. In our scenarios, these span across almost two orders of magnitude, as shown by the sample Probability Density Function (PDF) of $r$, $\lambda$ and $w$ shown in Figure~\ref{fig2}~(g-h-i), respectively. \\
	
	\section{Microfluidics experiments}
	
\begin{figure}[h!]
		\includegraphics[trim={0 0 0 0},clip,width=1\textwidth]{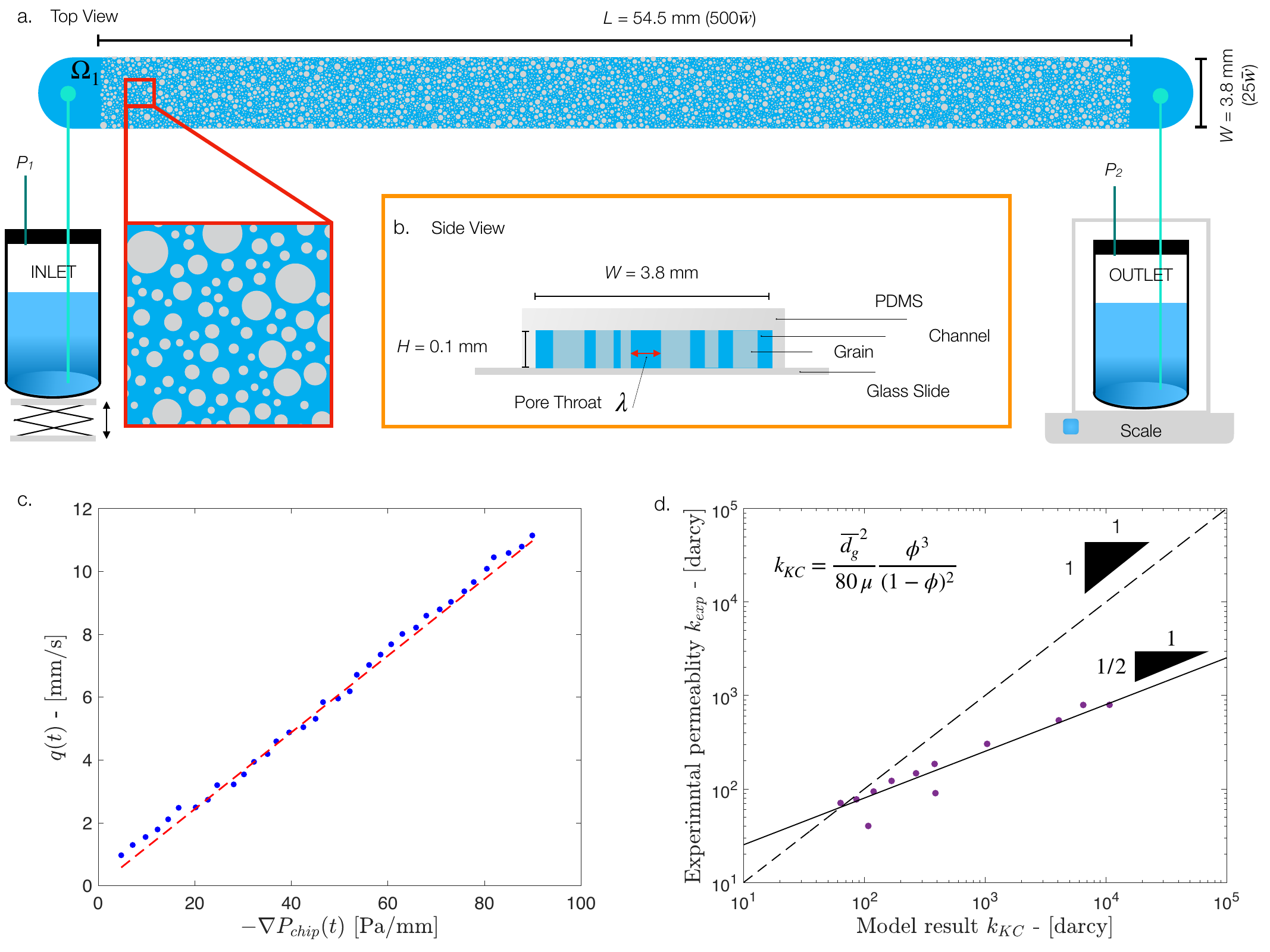} \\
		\caption{(a) Schematic view of the experimental set-up to measure the intrinsic permeability of the host medium by imposing a constant pressure gradient across a microfluidic device and continuously weighting the outlet reservoir to measure the associated flow rate. Gray disks correspond to grains, blue background indicates the water that saturates the pore space. The inlet reservoir is positioned on a laboratory jack to regulate water level and guarantee equal head between the two reservoirs and the chip at the beginning of the experiment. The outlet reservoir is placed on an analytical scale to periodically measure its weight. (b) Cross-section of the PDMS microfluidic device with pillars (gray) corresponding to grains, width $W = 3.8 $ mm and thickness $H = 0.1$ mm. It is plasma-bonded to a microscopy glass slide, the pore space being shown by the blue patches between the gray pillars. (c) Blue dots represent the Darcy velocity of the porous system under diverse pressure gradient across the chip. The red dashed line represents the best linear fit. (d) Experimental permeability $k_{exp}$ of the 12 heterogeneous media and prediction $k_{KC}$ of the classical Kozeny-Carman formula.}\label{fig3}
	\end{figure}

	\noindent
	We fabricate 12 porous microfluidic devices with the geometries described above and shown in Figure~\ref{fig3}~$a$. The solid matrix is composed by cylindrical pillars of vertical thickness $H = 100\, \mu$m (see Figure~\ref{fig3}~$c$), each mimicking a grain of the porous medium. The value of $H$ is selected so that it is similar to the average pore throat $\bar{\lambda}$. This avoids plug flow between grains characterized by a flat velocity profile among grains~\cite{de2021chemotaxis,bordoloi2022structure}. Fabrication of the microfluidics devices is based on standard soft-lithography~\cite{xia1998soft,bordoloi2022structure}. A silicon wafer is spin-coated with the SU-8~2025 permanent epoxy negative photo-resist (\textit{MicroChem}) to pattern a mold based on the designed geometries. Liquid polydimethylsiloxane (PDMS) is prepared as 1:10 by-weight mixture of Curing Agent and Base silicone elastomer (\textit{Sylgard} 184, \textit{Dow Europe Gmbh}). It is, then, poured onto the hard SU-8 mold. All PDMS chips are characterized by the same width $W = 3.8$~mm and length $L = 54.5$~mm. They are pinched for inlet and outlet holes and plasma bonded to a glass slide. To eliminate air bubbles that could be trapped during the water injection, the PDMS is degassed in a desiccator for about 30 minutes before each experiment~\cite{de2021chemotaxis,scheidweiler2024spatial}.\\
	
	\noindent
	The intrinsic permeability $k$ of the system is assessed experimentally by imposing a pressure drop across the porous domain and continuously measuring the corresponding flow rate $Q$ by weighting the outlet reservoir, as described below and shown schematically in Figure~\ref{fig3}~$a$ and $b$. Inlet and outlet reservoirs are 50~mL falcon tubes closed with gas-tight caps equipped with two threaded ports. One of the ports is connected to a pressure controller (OBI1 - MK3, \textit{ElveFlow}) through a tube with an inner diameter of~$3$~mm. The other port is connected to the microfluidics inlet/outlet with a Tygon tube with inner diameter of~$0.5$~mm. Pressure in each reservoir is controlled via the pressure controller. The outlet reservoir is placed on an analytical scale (XS205DU, \textit{Metzler-Toledo}) that continuously provides measurements of the reservoir weight to a computer, every 0.2~s. \\
	
	\noindent
	We perform flow experiments for each porous structure by continuously injecting milliQ-water into the previously saturated system. While the outlet pressure $P_2$ is set to the constant value of 10~mbar, we impose the inlet pressure $P_1$ to increase linearly with time from 10~mbar to 200~mbar during an interval of 1800~s.\\
	
	\noindent
	The imposed pressure drops are set to always result in flow rates ensuring laminar flow conditions across the pore space. In other words, our experiments are always characterized by a Reynolds number $ Re = \rho \bar{\lambda} q / \mu < 1$, where $\mu$ and $\rho$ are the dynamic viscosity and density of water at the experimental temperature conditions (21~C$^{\circ}$), respectively. The averaged (Darcy) velocity of the fluid $q=Q/A$ is computed from the measured flow rate $Q$ under the maximum pressure drop $\Delta P_t = P_2 - P_1 = - 190~mbar$ across the chip cross-sectional area $A = W H$). \\
	
	\noindent
	The time series $t_1$ and $t_2$ at which the inlet/outlet pressures $P_1(t_1)$, $P_2(t_1)$ and the outlet reservoir mass ($M(t_2)$) are measured differ as they correspond to data collected by different instruments. Thus, we define a common time $t$ represented by a series of $n = 100$ values from $t_0 = 0$~s and $t_n = 1800$~s, respectively corresponding to the experiment start, i.e. no flow condition $P_{out} = P_{in}$, and its end, when $ P_2 = P_1 - 190 ~  mbar$. Thus, the time interval $\Delta t$ of the common time is $1800/n = 18$~s. We then interpolate linearly pressure and mass data on such common time reference frame. The overall volumetric flow rate at each time, $Q(t)$, is evaluated form the interpolated out-flowing fluid mass, $M(t)$, as $Q(t) = \Delta M(t)/\rho/\Delta t$. Figure~\ref{fig3}~(c) depicts Darcy velocity $q$ as a function of the overall (temporally varying) pressure gradient. Finally, we calculate the intrinsic permeability $k_{exp}$ of the microfluidic devices by fitting the Darcy's law
	\begin{equation}\label{eqdacrylaw2}
		q(t) = - \frac{k_{exp}}{\mu} \, \nabla P_{c}(t),
	\end{equation}
	where $\nabla P_c(t) = \Delta P_c(t)/L_{c}$ is pressure gradient between the inlet and outlet of the chip. Beside pressure gradient variations that we impose via the pressure controller ($\Delta P_t = P_2 - P_1$), fluid pressure across the chip also varies during the flow experiments due to two mechanisms. \\
	
	\noindent
	First, we must consider variations in water level within the inlet and outlet reservoirs. The former decreases while the latter increases, as water flows from one reservoir to the other one. Then, pressure losses across the inlet and outlet connecting pipes should also be taken into account. Thus, the pressure drop across the whole device can be expressed as $ \Delta P_c(t) = \Delta P_t(t) + \Delta H(t) + \Delta P_l(t)$, where the last two terms represents the mentioned two mechanisms and are defined as follows. Since the inlet-outlet reservoirs have identical cross-sectional area $A_r$ and the water free surface therein is initially located at the same height, the inlet-outlet hydraulic head difference is estimated as twice the water level difference measured within one of the two reservoirs. The latter is equal to twice the volume of water flown since the beginning of the experiment divided by the reservoir cross-sectional area $A_r$, i.e., $\Delta H(t) = 2 (M(t) - M(0)) / \rho A_r$. The pressure loss $\Delta P_l$ associated with the fluid motion within the connecting pipes is given by the Hagen-Poiseuille law~\cite{sutera1993history} as $\Delta P_l(t) = 8\mu Q(t)L_p / \pi r_p^4$, where $L_p$ and $r_p$ correspond to the length and radius of the Tygon tube, respectively.\\
	
	\noindent
	The blue dots in Figure~\ref{fig3}~(c) represent the experimental data of the measured Darcy velocity $q(t)$ and the pressure gradient $\nabla P_c(t)$ across one of the porous structures designed ($\Omega_1$). Assuming that the data corresponding to Darcy velocity and pressure drop are affected by identically distributed small (about a few percentage) and random errors, we invoke the maximum likelihood principle to analytically calculate the best fit and estimate for the linear relationship coefficient, from the series of $n$ measured values of $q$ and $\nabla P_c$~\cite{taylor1982introduction}. Thus, we evaluate the intrinsic medium permeability $k_{exp}$ from the linear fit of eq.~\eqref{eqdacrylaw2} as:
	\begin{equation}\label{eq linear regression}
		k_{exp}= \frac{n \sum_{i=1}^{n} ((-\nabla P_{ci} / \mu) q_i) - \sum_{i=1}^{n} (-\nabla P_{ci} / \mu) \sum_{1}^{n} (q_i)}{n \sum_{i=1}^{n} (-\nabla P_{ci} / \mu)^2 - (\sum_{i=1}^{n}(-\nabla P_{ci} / \mu))^2},
	\end{equation}
	where its uncertainty is given by~\cite{taylor1982introduction},
	\begin{equation}\label{eq linear regression2}
		\delta_{k _{exp}}= \delta _{q} \sqrt{\frac{n}{n\sum_{i=1}^{n}(-\nabla P_{ci}/\mu)^2-(\sum_{i=1}^{n}(-\nabla P_{ci}/\mu))^2}}
	\end{equation}
	and
	\begin{equation}
		\delta_{q}=\sqrt{\frac{1}{n-2}\sum_{i=1}^{n}(q_i-\frac{n\sum_{i=1}^{n} ((-\nabla P_{ci}/\mu)^2 \sum_{i=1}^{n} q_i)-\sum_{i=1}^{n}(-\nabla P_{ci}/\mu)\sum_{i=1}^{n}((-\nabla P_{ci}/\mu)q_i)}{\sum_{i=1}^{n}(-\nabla P_{ci}/\mu)^2-(\sum_{i=1}^{n}(-\nabla P_{ci}/\mu))^2}-k_{exp}(-\nabla P_{ci}/\mu))^2}.
	\end{equation}

	\noindent
	The resulting permeability values are listed in Table~\ref{tab1}. These vary within the range $[40.2, \, 794]$~darcy (i.e., spanning more than one order of magnitude) with relative uncertainty $\delta_{k_{exp} }/ k_{exp}$ ranging between 0.2\% and 8\% (average value over the set of 12 structures being 2\%).\\
	\begin{table}[h]
		\centering
		\begin{tabular}{|l|cccccccccccc|}
			\toprule
			Structure & $\Omega_1$ & $\Omega_2$ & $\Omega_3$ & $\Omega_4$ & $\Omega_5$ & $\Omega_6$ & $\Omega_7$ & $\Omega_8$ & $\Omega_9$ & $\Omega_{10}$ & $\Omega_{11}$ & $\Omega_{12}$ \\ 
			\hline
			porosity $\phi$ [-] & 58.1\% & 63.9 \%& 71.6\% & 82.6\% & 85.8\% & 88.7\% & 54.5\% & 61.2\% & 51.4\%& 51.3\%& 51.6\% & 46.0\%\\ 
			\hline
			permeability $k_{exp}$ [darcy] &  122 & 185 & 303 & 542 & 792 & 794 & 77.4 & 147.2 & 71 & 94.1 & 90.2 & 40.2 \\
			\hline
			uncertainty $\delta _{k_{exp}}$ [darcy] & 1.5 & 3 & 8 & 16 & 40 & 68 & 0.5 & 2 & 0.2 & 0.6 & 0.7 & 0.2 \\
			\hline
		\end{tabular}
		\caption{Values of intrinsic permeability $k_{exp}$ and its related uncertainty $\delta_{k_{exp}}$ obtained for the 12 designed porous systems via linear interpolation of collected data~\cite{taylor1982introduction}. The relative uncertainty $\delta _{k_{exp}}/ k_{exp}$ results to range within the interval [0.2\%-8\%] with averaged value over the set of 12 structures being 2\%. }
		\label{tab1}
	\end{table}
	
	\noindent
	Figure~\ref{fig3}~(d) shows the experimentally based values of the 12~porous media permeability listed in table~\ref{tab1} against the classical Kozeny-Carman prediction, eq.~\eqref{eqKC}. As stated above, the latter provides an estimate of the medium permeability upon considering the resistance to flow as rendered by the Hagen Poiseuille law, accounting for the actual friction experienced by the fluid through an equivalent medium composed of identical grains of size $d_g$ equal to the average grain size (gray disks in Figure~\ref{fig1}). For an assemblage of cylinders, $c_0/\sigma = d_g^2 / 80$, which yields
	\begin{equation}\label{eqKC2}
		k_{KC} = \frac{d_g^2}{80 \, \mu} \frac{\phi^3}{(1-\phi)^2},
	\end{equation}
	where $d_g = \overline{r}$ is a mean grain diameter and $\phi$ is the medium porosity.\\
	
	\noindent
	This result does not provide an appropriate representation of the experimental data, both in terms of actual values and scaling pattern. The key reason for this discrepancy is that the Kozeny-Carman formulation does not explicitly takes into account the variability of the porous structures. In other words, the average grain size and porosity values are not representative of the structural variability of the medium, that should be taken into account.\\

	\section{Pore size variability control on porous medium permeability}
	
	\noindent
	To take into account the impact of the spatial variability of the pore space across the medium structure and upscale it to a consistent and sound assessment of the medium permeability, we consider the whole system as a collection of $m$ virtual porous systems in series. Each of these is characterized by a length $l_i (i=1,…,m)$ along the mean flow direction and a permeability $k_i$. The number of porous media of average length $\overline{l}$ needed to cover the system size is $m = L / \overline{l}$. In the following, we directly relate permeability $k_i$ and length $l_i$ to the actual medium structural properties. \\
	
\begin{figure}[h!]
		\includegraphics[trim={0 0 0 0},clip,width=1\textwidth]{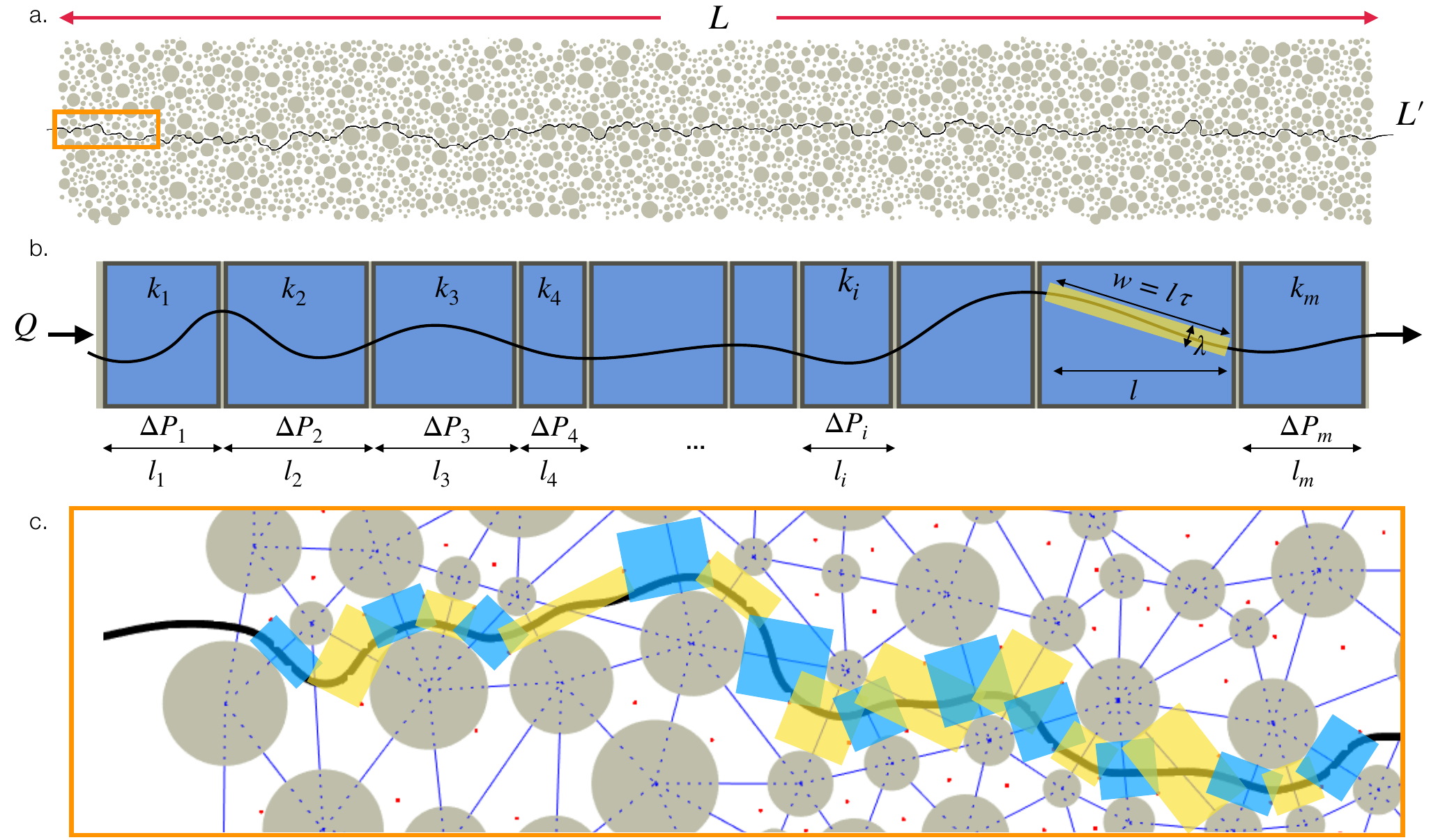} \\
		\caption{(a) Two-dimensional (2D) representation of the porous geometry $\Omega_1$ with a trajectory generated from particle tracking simulations through the numerically computed velocity field across the system. (b) Schematic view of our model, conceptualizing the system as a series of porous media of individual length $l_i$ and permeability $k_i$. (c) Zoomed-in view of the orange rectangle in (a); the trajectory moves across pores as they were pipes of diameter equal to the local pore throat and length equal to the local pore body.} \label{fig4}
	\end{figure}

	\noindent
	We start by considering the overall pressure drop $\Delta P_p$ experienced by a fluid particle $p$ flowing with average velocity $q$ along its trajectory of length $L^\prime = L \tau$, where $\tau = L^\prime / L$ represents the medium tortuosity, (as schematically shown in Figure~\ref{fig4}~(a).
	As the fluid particle moves across the series of $m$ smaller porous media, the pressure drop $\Delta P$ can be viewed as the sum of all pressure drops $\Delta P_i$ across each of individual medium of length $l_i$ (as in Figure~\ref{fig4}~(b)), i.e.,
	\begin{equation}\label{eq gradP3}
		\Delta P = - \frac{q \mu L \tau}{k} = - \sum_{i=1}^m q \mu \, \frac{l_i}{k_i}.
	\end{equation}
	The overall permeability $k$ is, then, estimated as the following harmonic average
	\begin{equation}\label{eq gradP4}
		k = \frac{L \, \tau}{ \sum_{i=1}^m \frac{l_i}{k_i} }.
	\end{equation} \\

	\noindent
	To relate the permeability $k_i$ and length $l_i$ of each of the $m$ porous media described above to the pore scale structural properties of the actual medium, we consider that the smallest portion of the macroscopic system that can host fluid flow is the individual pore. Thus, we assume that the value of permeability $k_i$ is the one of a pipe with diameter equal to a pore throat $\lambda$, i.e., $k_i = \lambda^2_i / 32$ (as in Figure~\ref{fig4}~(c)). The length $l_i$ of such a flow system equals the length of a pore body divided by the system tortuosity $l_i = w_i / \tau$, thus enabling one to take into account the random orientation of the local fluid flow direction with respect to the average flow (see Figure~\ref{fig4}~(b,c)). From a macroscopic perspective, the porous medium is, thus, conceptualized as composed by a series of $m = L \tau / \overline{w}$ individual porous systems, with overall permeability $k_t$ that depends on the microscopic pore throat $\lambda$ and body $w$ sizes according to
	\begin{equation}\label{eqpermtot}
		k_t = \frac{L \tau}{\sum_{i=1}^m \frac{l_i}{k_i}} = \frac{L \tau^2}{\sum_{i=1}^m \frac{32 \, w_i}{\lambda^2_i}}.
	\end{equation}

	\noindent
	Figure \ref{fig6}~(a) shows as green dots the scatter-plot of our measured permeability $k_{exp}$ and our model results (eq.~\eqref{eqpermtot}). It also includes the corresponding prediction based on the Kozeny-Carman formula (pink dots).
	We quantify the accuracy of our results across all designed 12 porous geometries in terms of the mean squared discrepancy $\chi$ between experimentally measured data $k_{exp}$ and model results $k_t$:
	\begin{equation}\label{chi}
		\chi = \sqrt {\frac{1}{N}\sum_{s=1}^{N} \left( \frac{k_{t_s} - k_{exp_s}}{k_{exp_s}} \right)^2}.
	\end{equation}
	While $\chi = 32 \%$ for our model, the Kozeny-Carman prediction yields a value of $\chi = 476\%$. \\
	
	\begin{figure}[h!]
		\includegraphics[trim={0 0 0 0},clip,width=1\textwidth]{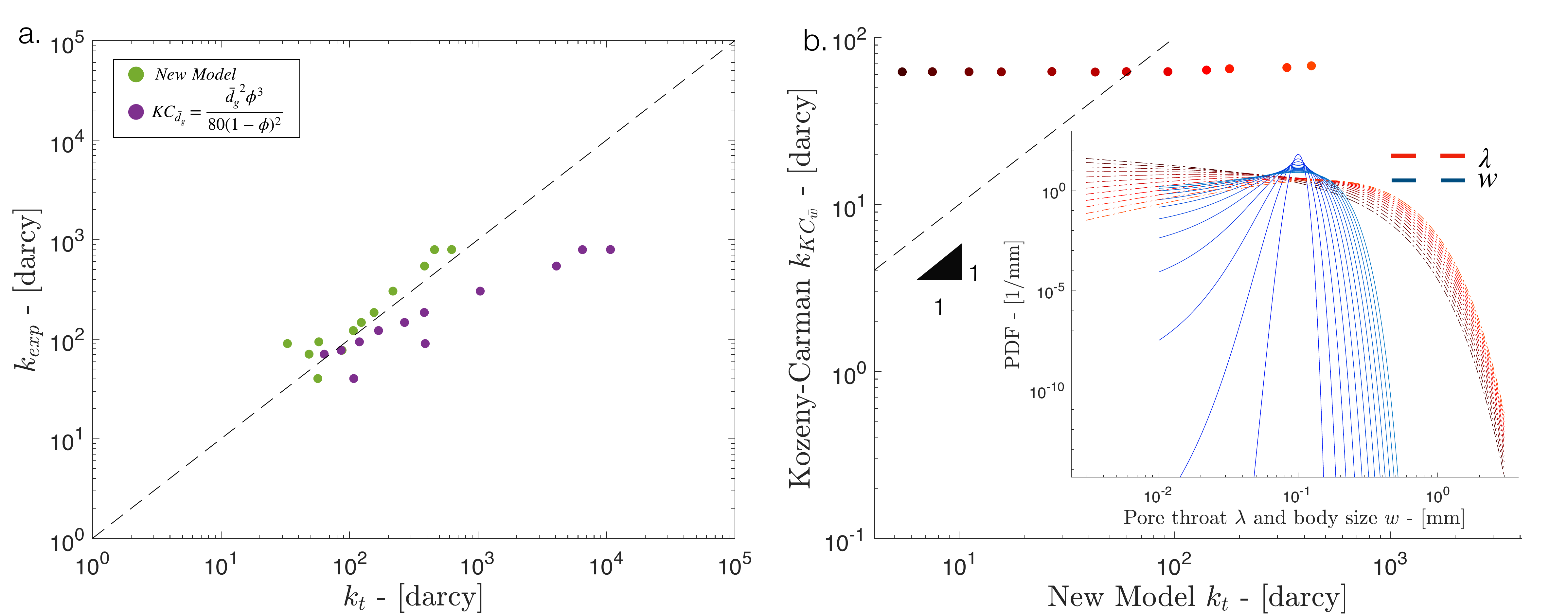} \\
		\caption{(a) Scatter-plot of experimental permeability measured for the heterogeneous media designed versus the prediction of ($i$) our model eq.~\eqref{eqpermtot} (green dots) or ($ii$) the Kozeny-Carman formula using the average grain diameter (pink dots) or average pore throat (pink stars) as characteristic length scale; (b) Scatter-plot of permeability values obtained through our model vs those obtained through the Kozeny-Carman formulation for 12 additional pore size broader distributions (shown in inset) characterized by th same averaged value.}\label{fig6}
	\end{figure}

	\noindent
	To test the robustness of our findings, we repeat the analysis upon defining pore sizes across the porous domains through a MIC (Maximum Inscribed Circles) algorithm~\cite{silin2006pore, gerke2020improving, bordoloi2022structure} (see Figure~\ref{fig6}~(a,b)). This approach relies centering a circle that touches the closest grain at each point of the pore space skeleton (see red disks in Figure~\ref{fig6}). We, then, consider each circle defined by the MIC algorithm to represent the individual pores, where the throat $\lambda_{mic}$ and body $w_{mic}$ sizes are both equal to the circle diameter $d_{MIC}$. Embedding this approach in our formulation yields a value of $\chi = 35 \%$, similar to the prediction obtained with the triangulation method, highlighting the robustness of our model. \\
	
	\noindent 
	We note that to ensure a parabolic fluid velocity profile between grains and avoid plug flow the microfluidics structures are limited by their own thickness. Thus, as discussed above, the ratio between largest and smallest size of throats or bodies is limited to around 80 (see Figure~\ref{fig2}~(h,i)). To further strengthen the relevance of our model, we then generate 12 additional virtual pore throat and body size distributions that are broader than what we could fabricate. These synthetic pore space distributions have all by the same porosity $\phi=0.5$, average pore throat $\bar{\lambda}= 0.09$~mm and pore body size $\bar{w} = 0.1$~mm. The corresponding distributions (as rendered through sample PDFs) of $w$ (blue dashed curves) and $\lambda$ (red dashed curves) are shown in the inset of Figure~\ref{fig6}~(b), where we depict a scatter plot of permeability values obtained through our model (eq. ~\eqref{eqpermtot}) versus those obtained through the Kozeny-Carman formula (eq.~\eqref{eqKC2}). While the latter predicts the same permeability for all the 12 geometries, since they have been designed to have the same macroscopic properties, our model taking into account the detailed pore size variability returns very different values.
	\begin{figure}[h!]
		\includegraphics[trim={0 0 0 0},clip,width=1\textwidth]{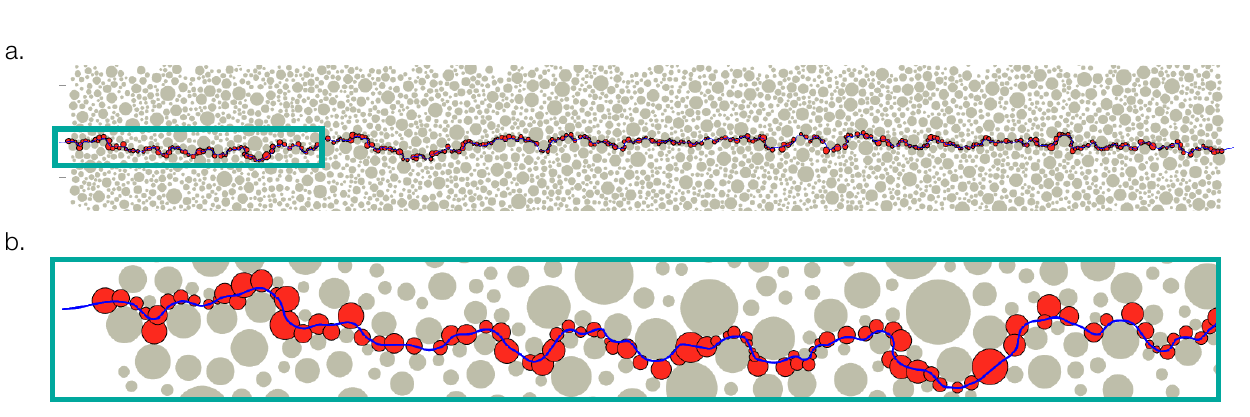} \\
		\caption{(a) The 2D porous geometry $\Omega_1$ with an exemplary trajectory (blue curve) generated from particle tracking based on a numerical solution of the velocity field. (b) Zoomed-in view of the green rectangle in (a); solid red circles (MICs) represents pores along the trajectory.}\label{fig6}
	\end{figure}

	\section{Conclusion}
	
	\noindent
	Despite the relevance of flow through porous media and related (flow-driven) processes, the dependence of a heterogeneous medium intrinsic permeability from its structural properties is still not fully understood. Here, we investigate this long-standing challenge by designing a set of heterogeneous porous structures that differ in the statistical distribution of their pore size. The latter span across more than one order of magnitude in our systems. We quantify their permeability with a novel experimental method by flowing water through microfluidics replicating a set of designed and complex porous structures, by imposing a controlled pressure gradient and continuously measuring the resulting fluid flow rate. \\
	
	\noindent
	The classical Kozeny-Carman formulation fails at rendering the permeability of all of the 12 structures analyzed, both in terms of values and scaling features. We then propose a new model that fully takes into account the structural variability of pore sizes and predicts the experimentally-based intrinsic permeability without the need of any fitting/tunable parameter. \\
	
	\noindent
	We incorporate the structural heterogeneity of the porous media analyzed by conceptualizing their macroscopic structure as composed by $m$ smaller porous media placed in series. We relate their number, $m$, individual length and permeability to the structural medium properties via the Hagen-Poiseuille law, describing flow through a pipe. We show that knowledge of the pore size $\lambda$ distribution is enough to accurately assess the overall intrinsic permeability for the scenarios that we analyzed. Our model fully embed the system tortuosity and establishes the hydraulic equivalence between the overall porous medium and a system of $m$ porous media in series with the same hydraulic properties of pipes with diameter $\lambda$. Hence, the medium permeability corresponds to the harmonic mean of $m$ individual permeability values $\lambda^2_i / 32$ ($i=1,…,m$) each weighted by the body size.

	\section*{ACKNOWLEDGEMENTS}
	\noindent 
	PdA acknowledges the support from the FET-Open project NARCISO (ID: 828890) and of Swiss National Science Foundation (grant ID 200021\_219863). WJ acknowledges the funding of China Scholarship Council for financial support through the fellowship (grant ID: CSC202008210309). AG acknowledges support from the European Union Next-Generation EU (National Recovery and Resilience Plan - NRRP, Mission 4, Component 2, Investment 1.3 - D.D. 1243 2/8/2022, PE0000005) in the context of the RETURN Extended Partnership. All authors thank Monica Riva, Martina Siena and Chiara Recalcati for useful and insightful discussions.\\

	\newpage

\end{document}